\documentclass[lettersize,journal]{IEEEtran}
\usepackage{colortbl} \usepackage{hyperref}
\definecolor{lightgray}{gray}{0.8}
\usepackage{amsmath,amsfonts}
\usepackage{amssymb}
\usepackage{booktabs}
\usepackage{multirow}
\usepackage{pifont} 
\usepackage{algorithm}
\usepackage{algpseudocode}
\usepackage{array}
\usepackage[caption=false,font=normalsize,labelfont=sf,textfont=sf]{subfig}
\usepackage{textcomp}
\usepackage{stfloats}
\usepackage{url}
\usepackage{verbatim}
\usepackage{graphicx}
 
\usepackage{xcolor}
\usepackage{microtype}
\usepackage{wrapfig}
\usepackage{lipsum}  
\usepackage{float}
\usepackage{siunitx}
\begin{document}

\title{ELEGANCE: Efficient LLM Guidance for Audio-Visual Target Speech Extraction}

\author{Wenxuan Wu, Shuai Wang$^\dagger$,~\IEEEmembership{Senior Member,~IEEE}, Xixin Wu,~\IEEEmembership{Member,~IEEE}, \\Helen Meng,~\IEEEmembership{Fellow,~IEEE}, Haizhou Li$^\dagger$,~\IEEEmembership{Fellow,~IEEE} 

\thanks{$^\dagger$: Corresponding author.}}




\maketitle

\begin{abstract}

Audio-visual target speaker extraction (AV-TSE) models primarily rely on visual cues from the target speaker. However, humans also leverage linguistic knowledge, such as syntactic constraints, next word prediction, and prior knowledge of conversation, to extract target speech. Inspired by this observation, we propose ELEGANCE, a novel framework that incorporates linguistic knowledge from large language models (LLMs) into AV-TSE models through three distinct guidance strategies: output linguistic constraints, intermediate linguistic prediction, and input linguistic prior. Comprehensive experiments with RoBERTa, Qwen3-0.6B, and Qwen3-4B on two AV-TSE backbones demonstrate the effectiveness of our approach. Significant improvements are observed in challenging scenarios, including visual cue impaired, unseen languages, target speaker switches, increased interfering speakers, and out-of-domain test set. Demo page: \href{https://alexwxwu.github.io/ELEGANCE/}{https://alexwxwu.github.io/ELEGANCE/}.
\end{abstract}


\section{Introduction}
Audio-visual target speaker extraction (AV-TSE) aims to isolate the target speaker's voice from a mixture speech signal by leveraging visual cues, particularly lip movements. This task mimics the human auditory system, seamlessly integrating lip movements and speech signals to focus attention on the voice of a single target speaker, providing insights into human brain function \cite{brain,select-listen}. 
AV-TSE is also crucial for enabling natural and intuitive interactions in many real-world scenarios. In human-robot interaction, AV-TSE allows a robot to focus on the commands of a specific user, even in a crowded environment \cite{HRI}. In online meetings, it can automatically isolate the active speaker, minimizing distractions and improving communication clarity \cite{misp}. Furthermore, AV-TSE serves as a vital front-end for the face-to-face spoken interaction model in the real world, enabling individuals with hearing impairments or those struggling to hear clearly to focus on conversations \cite{gpt4o}.

Inspired by the alignment mechanism in the human brain, many previous works on AV-TSE managed to better sync the audio and visual signals in temporal and feature space \cite{muse, av-sepformer,wu2024target_cvpr, ijcnn} and achieve significant extraction performance. To address scenarios where visual cues are compromised, researchers have developed several innovative approaches. Some studies introduce visual imagination mechanisms that associate target speaker lip movements with voice characteristics through additional visual recovery modules \cite{ImagineNET}. Others implement momentum mechanisms that recall historical extracted speech for continuous extraction when visual cues become unavailable \cite{li2025memo}. While these approaches align with human cocktail party experiences, challenges persist in scenarios involving unseen visual impairments and complex acoustic environments. 
More recently, researchers have explored mimicking the ``phoneme restoration effect" to handle imperfect extractions, $C^2$ AV-TSE 
 proposed a MAR strategy to leverage both inter and intra-modality context and a cascaded confidence model to automatically predict and refine those unclean extraction segments, potentially caused by
 visual cue impairment, more interfering speakers, or transient noise, etc \cite{wu2025c}.

Despite significant progress, the potential of linguistic knowledge remains largely unexplored in AV-TSE systems. According to speech chain theory, the language center facilitates the conversion of speech signals to semantic understanding \cite{avsepchain}. However, the utilization of deep semantics from text, beyond the audio-visual signal, remains underexplored for AV-TSE models. To this end, Large Language Models (LLMs), encoding rich linguistic knowledge, offer a promising source of external knowledge for AV-TSE. While \cite{wu2025incorporating} has pioneered the use of LLM-derived linguistic constraints for AV-TSE output, demonstrating that incorporating textual information can surpass simply using ASR with CTC loss. This approach primarily adds supervision based on linguistic rules and patterns for correction. While the corrected linguistic knowledge is injected into the extractor, these constraints do not participate directly during the extraction process. We know humans can predict the target speaker's incoming words during the extraction process, and sometimes the prior knowledge of the conversation may also facilitate the extraction process.

Building upon prior work using linguistic constraints from LLM as output guidance, this study investigates the potential of linguistic prediction and linguistic prior knowledge to further enhance AV-TSE models. For linguistic prediction, we propose directly informing the AV-TSE extractor with the next token prediction (NTP) loss derived from an LLM. In this strategy, the extracted target speech features from the AV-TSE extractor are jointly modeled with the intermediate features from the LLM to calculate the NTP loss, leveraging the LLM's strong generation capabilities \cite{shi2024lmfusion}. This NTP loss is then backpropagated to the AV-TSE extractor, serving as a direct predictive cue during the extraction process. For linguistic prior knowledge, except for the primary visual cues and the mixture speech signal, we propose utilizing LLM-derived linguistic features as additional input to the AV-TSE model. Before feeding these features into the AV-TSE extractor, we retrieve mutual information between the linguistic features extracted from the text and the mixture speech signal. This serves as a valuable linguistic prior guidance for the extractor. To verify the effectiveness and generalization of our three proposed LLM guidance strategies, we selected USEV and AV-Mamba as AV-TSE backbones, with AV-Mamba representing the first application of Mamba to the AV-TSE task.

The contributions could be summarized as follows:

\begin{itemize}
    \item We propose ELEGANCE, a comprehensive framework incorporating three distinct LLM guidance strategies (linguistic constraints, linguistic prediction, and linguistic prior) to enhance AV-TSE models during training. Systematic investigation across different LLMs and sizes (RoBERTa \cite{Liu2019RoBERTaAR}, Qwen3-0.6B \cite{qwen3technicalreport}, and Qwen3-4B \cite{qwen3technicalreport}) demonstrates significant improvements compared to baselines, highlighting the effectiveness of our approach.
    \item We investigate whether injected linguistic knowledge compensates for impaired visual cues and transfers across different languages. Models pre-trained on English monolingual mixtures are evaluated on four distinct language monolingual mixtures. Our experiments demonstrate that linguistic knowledge effectively compensates for visual impairments and transfers to cross-lingual settings, making the approach practical for low-resource languages.

    \item We assess the robustness of our strategies in three challenging real-world inference scenarios: increased interfering speakers, target speaker switching, and out-of-domain test sets. Our strategies consistently achieve improvements, indicating strong generalization capabilities without requiring additional training data or fine-tuning.
\end{itemize}

 \vspace{-5pt}
\section{Related Work}
\vspace{-5pt}

\subsection{Audio-Visual Target Speech Extraction}
Audio-visual TSE (AV-TSE) models normally utilize the target speaker’s lip movements as primary extraction cues \cite{muse, av-sepformer}. Unlike audio-only TSE models, which can be sensitive to the background noise or vocal variations from speech enrollment \cite{tse_summary}, the visual cues provide a direct and noise-robust pointer to the target speaker.

However, AV-TSE models often face challenges due to unreliable visual cues caused by factors such as object occlusion, poor lighting, and other environmental constraints. To address these issues, recent research has explored cognitively inspired mechanisms. For example, when visual cues are missing, some approaches attempt to ``imagine" or reconstruct the target visual cues using specialized recovery modules \cite{ImagineNET}. Other methods focus on leveraging rich inter- and intra-modality context from available audio and visual signals to compensate for temporarily lost visual information \cite{wu2024target_cvpr, ijcnn, wu2025c}. 

While these techniques enhance the robustness of AV-TSE models at the signal level, they share a fundamental limitation: they operate without any understanding of the linguistic content being spoken. This ``semantic deafness" prevents them from utilizing contextual, grammatical, or semantic information to aid the extraction process, which is a critical capability in human listening \cite{avsepchain}.

\subsection{Incorporating Textual Information into Target Speech Extraction}

To explore the potential of textual modality for AV-TSE models, some studies attempted to utilize text as extraction cues. These include using descriptive prompts about the speaker's vocal style or the content of their speech to guide extraction \cite{LISTEN-CHAT-REMIX, huo2025beyond}, or leveraging the transcript of the preceding conversation as a contextual prior \cite{kim2025contextual}. These studies align the textual prompt and mixture speech signal into the same semantic space via cross-modality modules and have shown that text could be an effective extraction cue. While effective, these methods often depend on having specific prior knowledge (a target speaker description or conversation history) that may not be available in many practical scenarios.
To reduce the reliance on text during inference, some studies on speech separation \cite {hsieh2024multimodal} and speech enhancement \cite{hung2025linguistic} injected textual linguistic knowledge into models only at the training stage via an extra cross-modality loss, which avoids collecting transcripts during inference in practice.

Within the AV-TSE task, the exploration of using linguistic guidance from text is nascent. The pioneering work of \cite{wu2025incorporating} was the first to demonstrate that textual embeddings from a Pre-trained Language Model (PLM) could assist an AV-TSE model. However, its contribution was limited to applying linguistic constraints on the model's output, functioning primarily as a post-supervision mechanism rather than being directly integrated into the extraction process. 

In this study, we build upon previous work by exploring how linguistic knowledge can be more deeply embedded as predictive or prior guidance during the extraction process itself.

\vspace{-5pt}
\subsection{Large Language Models for Speech Guidance}
Benefiting from their strong generative capabilities and the rich knowledge acquired from vast amounts of unlabeled text, Large Language Models (LLMs) have successfully guided and enhanced numerous speech-related downstream tasks. The integration of LLM guidance into speech models can be broadly categorized into two approaches: direct integration of the LLM with speech models or knowledge distillation from the LLM into the speech models.

\subsubsection{Direct Integration} This approach typically cascades a Pre-trained Speech Language Model (PSLM) with a text-based LLM. The PSLM (e.g., Whisper \cite{whisper}) typically serves as an audio front-end, converting speech into a feature sequence that the LLM then processes for a downstream task, such as speech recognition \cite{whisper-llama,mengicassp2025}, synthesis \cite{LLMttst5, DiffDSR}, translation \cite{Cappellazzo2024LargeLM}, and emotion recognition \cite{thimonier2025emosllm}. While LLMs serve as powerful back-ends and provide useful linguistic knowledge for downstream tasks, these integrated systems are often computationally intensive, making them less suitable for resource-constrained scenarios, especially for real-time processing on edge devices.

\subsubsection{Knowledge Distillation and Transfer} 
Rather than direct integration, some approaches employ LLMs as knowledge bases and utilize knowledge distillation or knowledge transfer techniques to incorporate LLM guidance for downstream speech tasks through tailored loss functions. 
For example, in \cite{DistillationASR}, the authors designed various knowledge distillation losses to inject linguistic knowledge from a decoder-only LLM into the speech recognition decoding process, achieving a lower error rate without increasing inference costs.
Similar methods have also been applied to the speech separation task, \cite{hsieh2024multimodal} designed a strict temporal-alignment loss between speech and transcripts,  requiring alignment information in advance.  In \cite{lin2025bridging, hung2025linguistic}, the author transfers LLM guidance to a speech enhancement model via an extra cross-attention layer, and the fusion layer will be kept during inference. 

In this study, we opt for knowledge transfer methods and design three efficient strategies: LLM guidance from linguistic constraints, linguistic predictions, and linguistic prior. All three strategies do not require additional parameters during inference.
 \vspace{-5pt}
\section{ELEGANCE: Efficient LLM Guidance Framework}
This section introduces ELEGANCE, our proposed framework for incorporating linguistic knowledge from LLMs into AV-TSE models. We present three distinct guidance strategies that leverage different aspects of linguistic knowledge: output linguistic constraints, intermediate linguistic prediction, and input linguistic prior knowledge. All strategies employ a ``plug-and-play'' paradigm where LLMs provide guidance during training but are removed during inference, ensuring computational efficiency.

 \vspace{-5pt}
\begin{figure*}[htbp]
 \centering
\includegraphics[scale=0.45]{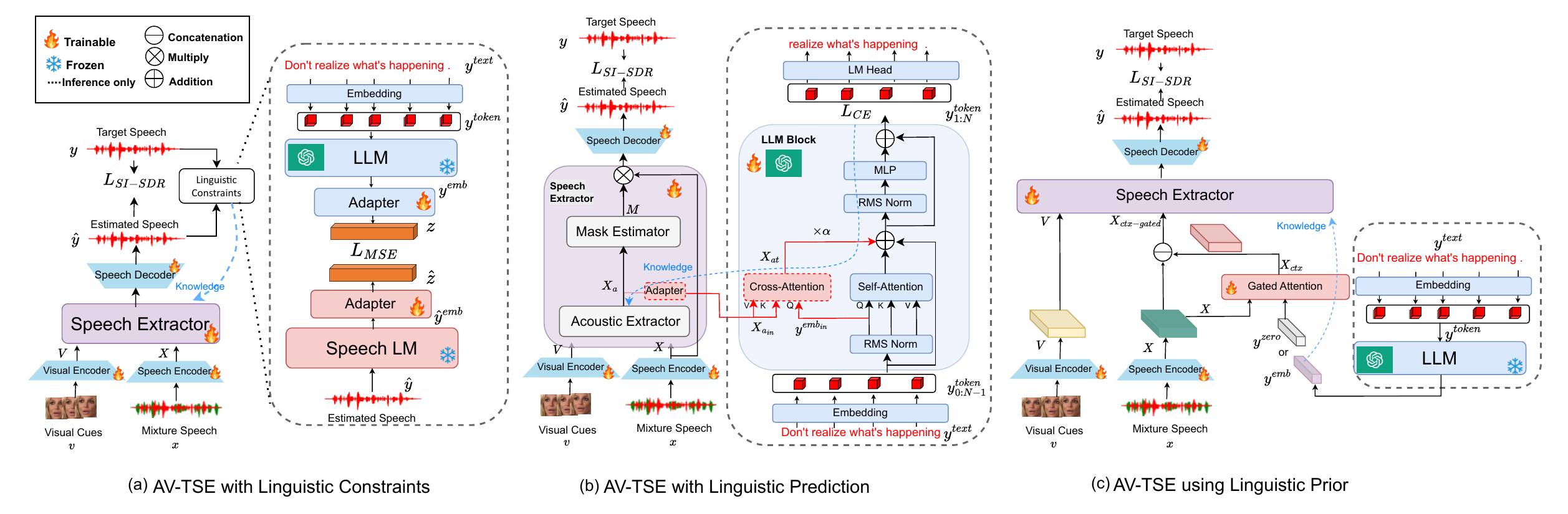}
\vspace{-20pt}
 \caption{\centering Three different LLM guidance strategies for AV-TSE, from left to right: (a) Output guidance: AV-TSE with linguistic constraints, where both PLM and PSLM will be utilized during training with an adapter to align latent semantic space. (b) Intermediate guidance: AV-TSE with Linguistic Prediction. Here, we use decoder-only LLM (Qwen) as an example; however, the strategy could also be applied to encoder-only LLMs, such as Roberta. By using causal architecture with a cascade transformer decoder, the fusion method will be the same.  (c) Input guidance: AV-TSE with Linguistic Prior. The LLM guidance and zero embedding will be added in an interleaved way. The goal is to reduce the over-reliance of the AV-TSE model on transcripts, which are not available during inference.
 Among all three strategies, the modules in the gray dotted lines will only be used during training and will be dropped during inference. The blue dotted lines denote the trajectory of linguistic knowledge injection. }
  \vspace{-5pt}
 \label{three_fusion}
\end{figure*}


\subsection{Three LLM Guidance Strategies}

In this section, we introduce three strategies designed to transfer linguistic knowledge from LLMs to AV-TSE models. All strategies employ a ``plug-in-plug-out'' training paradigm: LLMs are plugged in during training to provide guidance but are removed during inference, ensuring no additional computational overhead. Similar works on representation alignment for generation (REPA) have shown great success in accelerating convergence and improving generation quality in many diffusion models~\cite{yu2024repa}.

 In a previous study~\cite{wu2025incorporating}, we explored incorporating discrete or continuous linguistic constraints from PSLM as well as continuous linguistic constraints from PLM, and we found that using linguistic constraints from text is more effective~\cite{wu2025incorporating}. Thus, in this study, we focus on incorporating textual knowledge guidance from PLM for all three strategies. Specifically, we utilize Roberta-base, Qwen3-0.6B, and Qwen3-4B as knowledge bases. Among these, RoBERTa‑base is a non‑autoregressive (NAR) LLM pretrained with the Masking strategy, whereas Qwen3 is an autoregressive (AR) LLM with a more advanced architecture and pretrained with the Next-Token-Prediction (NTP) strategy. We introduce each guidance strategy as follows:

\subsubsection{Output Guidance: Linguistic Constraints}

For output guidance, we incorporate global linguistic constraints into the AV-TSE output by leveraging utterance-level semantic knowledge extracted from textual modality to guide the extraction process.  
Specifically, during training, we input target speech transcripts  $y^{text}$ into a frozen LLM, After the embedding layer,  $y^{text}$ will be discretized to $y^{token} \in R^{N}$, where $N$ is the text token sequence length. Then we input $y^{token}$ into LLM layers to derive the last layer embedding, represented as $y^{emb}$, where $y^{emb}\in R^{N\times C}$, where $C$ is the textual feature dimension. 
We then temporally pooled $y^{emb}$ to obtain utterance-level semantics $z$, where $z\in R^{C}$. 
Similarly, we input the predicted target speech $\hat{y}$ into a Pre-trained Speech Language Model (PSLM) and derive the last layer embedding, represented as $\hat{y}^{emb}$,   where $\hat{y}^{emb}\in R^{T\times C'}$, $T$ denotes the speech frame length and $C'$ is the speech feature dimension.  Then we temporally pooled $\hat{y}^{emb}$ to obtained $\hat{z}$, where $\hat{z}\in R^{C'}$. 

To align the latent semantic space between the textual feature and speech feature,  $z$ and $\hat{z}$ will be transformed by two adapter linear layers for channel projection. This process can be formulated as:
\begin{align}
z &= \text{Linear}(\text{Temporal Pooling}(y^{\text{emb}})), \quad \text{where } z \in \mathbb{R}^{D} \\
\hat{z} &= \text{Linear}(\text{Temporal Pooling}(\hat{y}^{\text{emb}})), \quad \text{where } \hat{z} \in \mathbb{R}^{D}
\end{align}
where $D$ denotes the aligned feature dimension. 

During training, both PLM and PSLM will be frozen. The symmetric adapter linear layers and the AV-TSE model will be fine-tuned. 
We use Mean Squared Error (MSE) to incorporate linguistic constraints from text transcripts to speech output.
The training objective $\theta$ is formulated as:
\begin{equation}
     \theta = \arg \min_{\theta} \mathcal{L}_{\text{SI-SDR}}(f_{\theta}(x, v), y) + \omega\mathcal{L}_{\text{MSE}}(\hat{z}, z),
\end{equation}
 where  $\omega$ controls the constraints scale.

\subsubsection{Intermediate Guidance: Linguistic Prediction}
For intermediate guidance, we enhance AV-TSE models by incorporating linguistic prediction knowledge, such as forecasting the target speaker’s next word. The linguistic prediction knowledge is encoded in LLM via the NTP training strategy, which serves as prediction cues for AV-TSE models. However, unlike global semantic knowledge, extracting such prediction knowledge from different LLMs and injecting it into different AV-TSE models could be challenging. On the one hand, most AV-TSE models are optimized offline, which is different from the NTP strategy. Besides, the AV-TSE model could be various and not limited to transformer-based architecture, and might be challenging to achieve a general interaction strategy leveraged in LMFusion \cite{shi2024lmfusion}.


To overcome such challenges and incorporate prediction cues into any AV-TSE models, we prefer to use a knowledge transfer strategy with NTP loss. LLM will be optimized with NTP loss, similar to the pretraining stage. Besides, for each vanilla LLM block, we further add a cross-attention layer. The LLM intermediate feature $y^{emb_{in}}$ serves as a query, and the target acoustic intermediate feature $X_{a_{in}}$  serves as a key and value, where $y^{emb_{in}} \in R^{N\times C_{in}}$ and $X_{a_{in}} \in R^{T\times C_{in}}$. The goal is to use LLM intermediate features to query similar patterns from the corresponding speaker's acoustic features, then retrieve mutual information $X_{at}$, where $X_{at} \in R^{ N \times C_{in}}$.

Note that $X_{a_{in}}$ is linear transformed by $X_{a}$, to align feature dimension of $y^{emb_{in}}$. Such a process could be formulated as:
\begin{equation}
X_{\text{at}} = \textbf{Cross-Attention}(y^{emb}, \text{Linear}(X_{a}), \text{Linear}(X_{a}))
\end{equation}

After the cross-attention operation, we assume $X_{at}$ has queried useful mutual information between target text and acoustic features. To make it play roles in the LLM modeling process,  we add $X_{at}$ as a residual to the vanilla LLM branch. To prevent excessive acoustic information from biasing the LLM's NTP process, we gate $X_{at}$ with a scale factor $\alpha$.

As shown in Fig. \ref{three_fusion}, the joint modeling of the AV-TSE task and the acoustic-informed LLM NTP task allows linguistic prediction knowledge to transfer back to the speech extractor through the cross-entropy (CE) loss, providing additional guidance. This design is motivated by two considerations:
\textbf{(1) Leveraging Linguistic Knowledge}: By connecting the AV-TSE model to the LLM through intermediate layers, the acoustic extractor can incorporate linguistic prediction knowledge from the NTP loss, complementing the reliance on visual cues. This approach mirrors how humans use both visual and linguistic cues in complex auditory environments.
\textbf{(2) Handling Extraction Errors}: For samples with inaccurate or noisy extraction, inconsistencies between the extracted acoustic features $X_{a_{in}}$ and the linguistic embeddings $y^{emb_{in}}$ can lead to uninformative or misleading features $X_{at}$. In such cases, the LLM’s NTP capability allows the CE loss to dominate and propagate backward, helping to refine the extraction process.

To balance the AV-TSE and LLM NTP tasks, we fine-tune all parameters of LLM for the first two epochs to boost convergence. Then, to avoid overfitting, only the cross-attention layer in each LLM block, the linear adapters, and AV-TSE models will be further fine-tuned.  The vanilla LLM parameters will be frozen. The training objective $\theta$ is formulated as:
\begin{equation}
\begin{aligned}
\label{NTP}
\theta &= \arg \min_{\theta}  \mathcal{L}_{\text{SI-SDR}}(f_{\theta}(x, v), y)
+ \delta \mathcal{L}_{\text{LM}}  , \\
\mathcal{L}_{\text{LM}} &= \mathbb{E}_{y^{\text{token}}} \left[ -\log P(y^{\text{token}}_i | y^{\text{token}}_{<i}, X_{at}) \right],
\end{aligned}
\end{equation}
As shown in Equation~\ref{NTP}, LLM will be optimized to predict the next text token $y^{\text{token}}_{i}$, conditioned on previous predicted text tokens $y^{\text{token}}_{<i}$, and acoustic-text mutual information $X_{at}$, and $\delta$ is used to control the NTP loss scale.


\subsubsection{Input Guidance: Linguistic Prior}
For LLM input guidance, we incorporate linguistic prior knowledge for the AV-TSE input, as shown in Fig. \ref{three_fusion},  except primary visual cues that enable ``seeing to hear better", prior knowledge on the target speech content typically facilitates easier extraction, enabling ``understand to hear better". Similar to the output guidance strategy, we derive the utterance-level LLM embedding $y^{emb}$ from the frozen LLM to provide global semantic knowledge. 

We then input $y^{emb}$ into the AV-TSE model as additional extraction cues to retrieve global mutual information between LLM embedding $y^{emb}$ and mixture speech feature embedding $X$.  Gated attention has been proven effective for dynamically obtaining global semantics between text and speech \cite{huo2025beyond}. Thus, we employ gated attention between two different modalities, where the mixture speech embedding $X$ serves as query, LLM embedding $y^{emb}$ serves as key and value, aiming to retrieve target speech content from the mixture. 
The gated attention is formulated as follows:
\begin{equation}
\label{gatedatt-attention}
\begin{aligned}
X_{\text{ctx}} &= \textbf{AttPooling}(X), \\
X_{\text{ctx}} &= \textbf{Linear}(\text{ReLU}(\text{Norm}(X_{\text{ctx}}))), \\
X_{\text{ctx}} &= \textbf{GatedUnit}(X_{\text{ctx}}, y^{\text{emb}}), \\
X_{\text{ctx-gated}} &= \textbf{Linear}(\text{Upsample}(X_{\text{ctx}}), X),
\end{aligned}
\end{equation}

where $y^{emb} \in \mathbb{R}^{1 \times C}$, $X_{ctx-gated} \in \mathbb{R}^{T \times D'}$, $C$ is the textual feature dimension, $D'$ is the gated speech feature dimension, and $T$ is the mixture feature frame length.

As shown in Equation~\ref{gatedatt-attention}, the mixture speech embedding $X$ will be passed into an attention pooling layer, norm layer, ReLU, and linear layer, respectively, to obtain contextual embedding $X_{ctx}$. To dynamically retrieve mutual information, $X_{ctx}$ will first be gated with $y^{emb}$ via a gated unit, then it will be upsampled temporally and concatenated with vanilla mixture embedding $X$. The concatenation result will be fed into a linear layer to obtain the gated mixture embedding $X_{ctx-gated}$. The preliminary experiments have shown that the concatenation is critical to extraction performance. Without concatenation with $X$, the contextual embedding may lose acoustic details and fail to predict the correct target speech mask.

Also, to prevent the model's over-reliance on $y^{text}$, which are typically unavailable in real-world scenarios, we employ a dropout strategy to reduce such reliance. Specifically, with probability $p$, we use the LLM embedding $y^{emb}$ as an additional input; with probability $1-p$, we replace $y^{emb}$ with a zero-embedding. The training objective $\theta$ is formulated as:

 \begin{equation}
\theta = 
\begin{cases}
\arg \min_{\theta} \mathcal{L}_{\text{SI-SDR}}(f_{\theta}(x, v, y^{emb}), y), & \text{with prob= } p, \\
\arg \min_{\theta} \mathcal{L}_{\text{SI-SDR}}(f_{\theta}(x, v), y), & \text{with prob=} 1 - p.
\end{cases}
\end{equation}
With proposed training objective, the AV-TSE models will be optimized with versatile input to enhance the sensitivity to textual information.

\section{EXPERIMENTAL SETUP}
\subsection{Baselines}
In this paper, we employ two different AV-TSE models as our baselines. The introductions of the two models could be summarized as follows:
\subsubsection{USEV} 
USEV is a time-domain AV-TSE model, which utilizes Dual-Path RNN (DP-RNN) blocks~\cite{luo2020dual} as the speaker extraction module.  Considering its training efficiency and lightweight nature, USEV has been widely used as a baseline in many previous studies \cite{usev}.  We also chose USEV as our primary baseline in this study.
\subsubsection{AV-Mamba}
Mamba is a lightweight state space selective model that offers linear computational complexity, in contrast to the quadratic complexity of transformers. Mamba-based speech models have demonstrated strong generalization on many downstream speech tasks, including speech enhancement \cite{chao2024investigation, kim25q_interspeech} and speech recognition \cite{Masuyama2024SLT}. To take advantage of both computation efficiency and strong generalization ability, we introduce AV-Mamba, the first Mamba-based time-domain AV-TSE model. AV-Mamba is a modified version of Mamba-TasNet \cite{jiang2025speech}, where the speech cue encoder is replaced by a visual cue encoder, and the Dual-Path Bidirectional-Mamba (DP-Bi-Mamba) blocks are utilized for the speaker extractor. AV-Mamba is designed to perform bidirectional scanning of the mixture speech, focusing on extracting key information relevant to the target speaker.

\subsection{Dataset}
In this study, several experimental settings are considered:
\subsubsection{Training and Validation Set}
We build two types of training sets: core training sets and English-only monolingual training sets:
\textbf{Core:} A two-speaker mixture training set is simulated following prior works~\cite{muse, ImagineNET, av-sepformer, wu2024target_cvpr}. The training set includes 20,000 utterances from 800 speakers. The SNR levels of interfering utterances are randomly sampled from -10 dB to 10 dB. Each mixture speech may involve two utterances from either different languages or the same language, selected at random. For the core validation set, it keeps the same setting but with 5000 utterances from 800 speakers in the training set.\\
\textbf{English monolingual:} Except core training set, we also simulated a two-speaker mixture training set, where both the target speaker and the interfering speaker speak English. In this setting, we aim to verify whether our LLM guidance strategies are language-dependent. The number of training and validation set utterances, as well as utterance durations, remains the same as the core training set.

\subsubsection{Test Set} 
Multiple test sets are designed to evaluate the proposed methods under diverse conditions:
    \textbf{Core:}  
    A two-speaker mixture test set is generated following methods from~\cite{muse, ImagineNET, av-sepformer, wu2024target_cvpr}, involving 118 randomly selected speakers and a total of 3,000 simulated mixtures. These mixtures include both same-language and cross-language pairs. During inference, all utterances are standardized to 6 seconds in length, with shorter utterances padded with zeros.

\begin{figure}[htbp]
 \centering
 \label{vocc}
\includegraphics[scale=0.41]{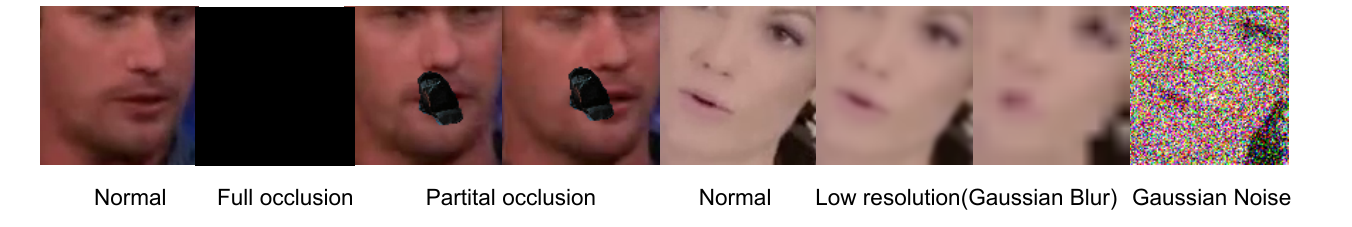}
\vspace{-15pt}
 \caption{\centering Examples of visual normal and impairment scenarios.}
\end{figure}

\textbf{Visual Cue Impaired:}  
Three visual impairment scenarios are simulated: partial occlusion, low resolution, and complete absence of visual information. The first two follow~\cite{hong2023watch}, where a randomly selected obstacle is used for partial occlusion, and Gaussian blur or noise is applied to simulate low-resolution conditions. Fully missing frames are replaced with zero values. Fig.\ref{vocc} presents examples of three visual impairment scenarios. For each scenario, the impairment ratio varies uniformly from 0\% to 100\%. All three visual impairment scenarios are selected with equal probability. The utterance numbers and durations remain the same with the core test set. 

   \textbf{Monolingual:}  
    Five two-speaker test sets are created for Italian (IT), Portuguese (PT), English (EN), Spanish (ES), and French (FR), with each mixture consisting of speakers from the same language. Due to limited test data for minority languages in VoxCeleb2~\cite{voxceleb2}, 118 speakers are randomly selected from the development set to generate 2,000 monolingual mixtures per language. The utterance numbers and durations remain the same with the core test set.

 \vspace{-7pt}

\textbf{Switching:} We simulated a target speaker switching scenario, where one interfering speaker remains active throughout, but the target speaker is changed to a different individual at specified intervals. The objective is to verify whether the AV-TSE model could seamlessly track and continuously extract different target speakers during the process. Specifically, the entire utterance is at least 16 seconds long, and the switch point, which marks the transition between two speech segments, is randomly sampled between 6-8 seconds. We also set the visual impairment ratio ranging from 0\% to 100\%. The visual impairment types are the same as in Fig.\ref{vocc}. The interfering speech is also mixed with a random SNR between -10 dB and 10 dB.

\textbf{Cross-Domain:}  
    We simulate an out-of-domain test set using the Lip Reading Sentences 3 (LRS3) test set~\cite{Afouras2018LRS3TEDAL}, which has been widely utilized in many audio-visual tasks~\cite{yeo2024akvsr, MultiModalCorrelation, deepavsr}
    For LRS3, we follow the same settings as the core test set.

\subsection{Metrics}
For evaluation, we employ several metrics. We use the scale-invariant source-to-distortion ratio (SI-SDR) \cite{SDR}, the improvement in SI-SDR (SI-SDR-i), and the improvement of signal-to-noise ratio (SDR-i) \cite{SDR} to measure signal similarity. To evaluate perceptual clarity and intelligibility, we also utilize the Perceptual Evaluation of Speech Quality (PESQ) \cite{pesq} and the Short-Term Objective Intelligibility (STOI) \cite{stoi}. To evaluate semantic similarity, we utilize SpeechBERTScore \cite{saeki2024speechbertscore}. For all metrics, a higher value indicates better performance.

\subsection{Implementation Details}
\subsubsection{AV-TSE Backbones}


For USEV, the Conv1D-based speech encoder and speech decoder have a hidden dimension of 256, with a kernel size of 40 and a stride of 20.  
For the DP-RNN-based speech extractor, the output channels are 64, the hidden channels are 128, the DP-RNN blocks are set to 6, and the chunk length is 100.

For AV-Mamba, the Conv1D-based speech encoder and speech decoder have a hidden dimension of 512, with a kernel size of 40 and a stride of 20.  
For the DP-Mamba-based speech extractor, the number of DP-Mamba blocks is set to 16, the Mamba bottleneck dimension is 512, and the Mamba state dimension is 16.  
The first linear projection has a factor of 2, and the kernel size of the Mamba convolution is 4.  
We set the residual precision to float32 to ensure training stability.

\subsubsection{LLM guidance strategies}

For output guidance, the dimensions of speech feature $C'$, and aligned feature $D$ are all set to $768$, the textual feature $C$ is set to $768$, $1024$, and $2560$ for Roberta-base, Qwen3-0.6B and Qwen3-4B, respectively.

For intermediate guidance, each cross-attention layer consists of $1$ heads, with a hidden dimension $D$ of $768$, $1024$, and $2560$ for Roberta-base, Qwen3-0.6B and Qwen3-4B, respectively, and Roberta-base is set to decoder mode. The scale factor $\alpha$ is set to $0.1$, preliminary experiments have shown that larger scale factors may lead to early stopping.  

For input guidance, the gated speech feature dimension $D'$ is set to $256$ and $512$ for USEV and AV-Mamba, respectively. the dropout rate probability $p$ is empirically set to $0.2$.

\subsubsection{Training Settings}
For input guidance and output guidance, the AV-TSE models are trained from scratch. For intermediate guidance, to prevent NTP modeling bias, we pre-trained the AV-TSE model and fine-tuned it with LLM guidance. The entire fine-tuning process is set to 30 epochs.
The learning rate is set to 1e-3 for input and output guidance, and to 15e-5 for intermediate guidance. During both pre-training and fine-tuning, the learning rate is halved when the validation loss decreases for 6 consecutive epochs, and training stops when the validation loss has not decreased for 10 epochs.

\begin{table*}[t]
  \caption{Offline inference on the core test set with clean visual cues. Both USEV and AV-Mamba are pre-trained on the core training set under the same clean-visual setup. “Guidance” denotes whether LLM guidance is applied at the Input, Intermediate, or Output stage of the pipeline (same LLM: RoBERTa-base). Bold numbers mark the best within each backbone (USEV or AV-Mamba).}
  \label{tab:diff_avtse_same_LLM}
  \centering
  \setlength{\tabcolsep}{6pt}
  \renewcommand{\arraystretch}{1.15}
  \begin{tabular}{
    l
    l
    S[table-format=2.3] 
    S[table-format=2.3] 
    S[table-format=2.3] 
    S[table-format=1.3] 
    S[table-format=1.3] 
    S[table-format=1.3] 
  }
    \toprule
    \multirow{2}{*}{\textbf{AV-TSE}} & \multirow{2}{*}{\textbf{Guidance}}
    & \multicolumn{3}{c}{\textbf{Distortion (dB)}} & \multicolumn{3}{c}{\textbf{Perceptual/Intelligibility/Semantic}} \\
    \cmidrule(lr){3-5} \cmidrule(lr){6-8}
    & & {\textbf{SI-SDR}} & {\textbf{SI-SDR-i}} & {\textbf{SDR-i}} & {\textbf{PESQ}} & {\textbf{STOI}} & {\textbf{SpeechBERTScore}} \\
    \midrule
 
    \multirow{4}{*}{USEV}
      & \cellcolor{lightgray}Baseline        & \cellcolor{lightgray}10.749 & \cellcolor{lightgray}10.797 & \cellcolor{lightgray}11.298 & \cellcolor{lightgray}2.845 & \cellcolor{lightgray}0.861 & \cellcolor{lightgray}0.806 \\
      & Input           & \bfseries 11.478 & \bfseries 11.525 & \bfseries 11.832 & 2.083 & \bfseries 0.869 & \bfseries 0.813 \\
      & Intermediate    & 11.297 & 11.344 & 11.652 & 2.847 & 0.867 & 0.810 \\
      & Output          & 11.416 & 11.464 & 11.777 & \bfseries 2.907 & 0.868 & 0.812 \\
    \midrule
 
    \multirow{4}{*}{AV-Mamba}
      &\cellcolor{lightgray}Baseline        & \cellcolor{lightgray}11.596 & \cellcolor{lightgray}11.643 & \cellcolor{lightgray}12.007 & \cellcolor{lightgray}2.965 & \cellcolor{lightgray}0.867 & \cellcolor{lightgray}0.817 \\
      & Input           & 11.611 & 11.658 & 12.051 & 2.931 & 0.871 & 0.827 \\
      & Intermediate    & \bfseries 12.673 & \bfseries 12.720 & \bfseries 13.053 & \bfseries 3.097 & \bfseries 0.890 & \bfseries 0.847 \\
      & Output          & 12.390 & 12.437 & 12.773 & 3.060 & 0.886 & 0.842 \\
    \bottomrule
  \end{tabular}
  \vspace{2pt}
 
\end{table*}

\section{Experimental Results}
\subsection{Do all three LLM guidance strategies generalize well to AV-TSE models? Which strategy is more effective?}



In this setting, we aim to verify the effectiveness of three proposed LLM guidance strategies on different AV-TSE models. To achieve this, we utilize USEV and AV-Mamba as AV-TSE backbones and choose Roberta-base as the knowledge base.

As shown in Table \ref{tab:diff_avtse_same_LLM}, we report evaluation results of baseline,  baseline with input guidance,  baseline with intermediate guidance, as well as baseline with output guidance for both USEV and AV-Mamba. Note that all models are pre-trained on the core training set and evaluated on the core test set. Both training and evaluation use clean visual cues.

We observe that for both USEV and AV-Mamba, all three strategies can improve performance in terms of signal similarity, speech intelligibility, and semantic similarity. Specifically, for USEV, input guidance yields the most significant improvement in SI-SDR, increasing from 10.749 to 11.478. For AV-Mamba, intermediate guidance produces the largest gain, with SI-SDR improving from 11.611 to 12.673. Both settings also achieve the best results in speech intelligibility and semantic similarity.

These findings suggest that the optimal LLM guidance strategy may vary depending on the AV-TSE backbone, likely due to different speech extraction modules favoring distinct linguistic cues. Additionally, we found that, for both USEV and AV-Mamba, using the input guidance strategy results in worse overall performance compared to the baseline, despite other metrics, such as SpeechBERTScore and STOI, still showing improvements with the proposed strategies. To explain this, we hypothesize that the input guidance strategy, which uses dropout to incorporate linguistic prior information into AV-TSE models, introduces differences in the input features compared to vanilla AV-TSE models; however, the model architecture is unchanged. This may lead the models to place less emphasis on the quality of speech reconstruction.

\subsection{Do LLM guidance provide more robustness to  AV-TSE models on unseen scenarios?}

\vspace{-5pt}
\begin{table*}
\caption{Cross-domain evaluation on LRS3 test set. We report USEV and AV-Mamba results using three guidance from Roberta. All results are OFFLINE INFERENCE with clean visual cues. }
\label{LRS3}
\centering
\setlength{\tabcolsep}{6pt} 
\renewcommand{\arraystretch}{1.15} 
\begin{tabular}{
    l 
    l 
    S[table-format=2.3] 
    S[table-format=2.3] 
    S[table-format=2.3] 
    S[table-format=1.3] 
    S[table-format=1.3] 
    S[table-format=1.3] 
  }
  \toprule
  \multirow{2}{*}{\textbf{AV-TSE}} & \multirow{2}{*}{\textbf{Guidance}}
  & \multicolumn{3}{c}{\textbf{Distortion (dB)}} & \multicolumn{3}{c}{\textbf{Perceptual/Intelligibility/Semantic}} \\
  \cmidrule(lr){3-5} \cmidrule(lr){6-8}
  & & \textbf{SI-SDR} & \textbf{SI-SDR-i} & \textbf{SDR-i} & \textbf{PESQ} & \textbf{STOI} & \textbf{SpeechBERTScore} \\
  \midrule
\multirow{3}{*}{USEV}
&  \cellcolor{lightgray}Baseline & \cellcolor{lightgray}12.041 & \cellcolor{lightgray}12.039 & \cellcolor{lightgray}12.465&\cellcolor{lightgray}2.821 & \cellcolor{lightgray}0.900 & \cellcolor{lightgray}0.826\\
  & Input & 12.722& 12.720 & 13.050& 1.802& 0.906&0.832  \\
 & Intermediate & 12.703  & 12.702&  13.009& 2.821& 0.908&0.833  \\
 &  Output &\textbf{12.816} & \textbf{12.815}& \textbf{13.121}& \textbf{2.891}&\textbf{0.909}&\textbf{0.837}\\
\hline
\multirow{3}{*}{AV-Mamba}
&  \cellcolor{lightgray}Baseline & \cellcolor{lightgray} 13.346 & \cellcolor{lightgray} 13.344 & \cellcolor{lightgray} 11.140&\cellcolor{lightgray}3.258 & \cellcolor{lightgray}0.9458 & \cellcolor{lightgray}0.9047\\
  & Input & 13.6770&13.67574 & 11.57977& 3.241690& 0.948400&0.89902  \\
 & Intermediate &\textbf{14.507}  & \textbf{14.506}& \textbf{12.353}& \textbf{3.408}& \textbf{0.960}&\textbf{0.921} \\
 &  Output &14.224& 14.228& 12.04468& 3.3563&0.95645&0.91528\\
 \toprule
\end{tabular}
\end{table*}

In this section, we aim to explore the generalization ability of the proposed LLM guidance strategies on unseen scenarios. We consider cases including target speaker switching during an utterance, interference from two simultaneous speakers, and evaluation on the out-of-domain LRS3 test set. Note that for all three scenarios,  we evaluate our pretrained models on the corresponding test set directly, without fine-tuning.
\subsubsection{Target Speaker Switching}
In this scenario, once the target speaker switches, the context of the target speech utterance changes immediately, posing a challenge to seamless extraction. As shown in Table     \ref{switch_res}, we report USEV and AV-Mamba performance on visual cue clean and impaired conditions with three strategies. Note that for visual cue impaired evaluation, we also use models pre-trained on the clean visual cue training set for direct comparison. We observe that for USEV and AV-Mamba, using input guidance and intermediate guidance achieves the best performance, respectively. AV-Mamba achieves more than 2 dB improvements on visual cue impaired settings.

\begin{table}
\centering
\caption{Offline inference \textbf{SI-SDR} results on target speaker switching test set. Note that all models are pre-trained on the core training set with clean visual cues. For all three LLM guidance strategies, we use Roberta.  We evaluate all pre-trained models on visual clean and visual impaired test set without additional fine-tuning.}
\setlength{\tabcolsep}{6pt}
\renewcommand{\arraystretch}{1.15}
\begin{tabular}{
    p{0.06\textwidth} 
    p{0.06\textwidth} 
    S[table-format=2.3] 
    S[table-format=2.3] 
    S[table-format=2.3] 
    S[table-format=2.3] 
  }
  \toprule
  \multirow{2}{*}{\textbf{Model}} & \multirow{2}{*}{\textbf{Visual cue}} & \multicolumn{4}{c}{\textbf{SI-SDR (dB)}} \\
  \cmidrule(lr){3-6}
  & & \textbf{Baseline} & \textbf{Input} & \textbf{Intermediate} & \textbf{Output} \\
  \midrule
\multirow{2}{*}{USEV} & Impaired & 8.971 & \textbf{9.447} & 8.975 & 9.183 \\
                      & Clean    & 9.853 & \textbf{10.286} & 9.929 & 9.786 \\
\midrule
\multirow{2}{*}{AV-Mamba} & Impaired & 9.075 & 9.529 & \textbf{12.167} & 10.269 \\
                          & Clean    & 10.739 & 11.258 & \textbf{12.439} & 11.655 \\
\bottomrule
\end{tabular}
\label{switch_res}
\end{table}

\begin{table}[!ht]
\begin{center}
\caption{OFFLINE INFERENCE with three-speaker mixture test set in zero-shot setting. We treat the extraction result with an SI-SNRi lower than 0 dB as a false extraction sample; the `False Rate' is calculated over the entire test set.}
\label{3spk}
\setlength{\tabcolsep}{6pt} 
\renewcommand{\arraystretch}{1.15} %
\begin{tabular}{p{0.06\textwidth}p{0.08\textwidth}p{0.1\textwidth}p{0.1\textwidth}}  
\toprule
\textbf{AV-TSE} & \textbf{Guidance} & \textbf{SI-SDR-i$(\uparrow)$}& \textbf{False Rate$(\downarrow)$}\\
\toprule
\multirow{4}{*}{USEV} &\cellcolor{lightgray}Baseline& \cellcolor{lightgray}6.943&\cellcolor{lightgray}6.34\% \\
                  & Input & 6.951 &5.73\%   \\
                 & Intermediate&  6.980&5.52\%  \\
                  & Output&  7.177 &5.10\%  \\
\hline
\multirow{4}{*}{AV-Mamba} &\cellcolor{lightgray}Baseline&  \cellcolor{lightgray}6.781&\cellcolor{lightgray} 9.90\%\\
                  & Input & 7.063  & 7.36\%  \\
                 & Intermediate& 7.352  & 7.16\%  \\
                  & Output& 7.340  & 6.86\%  \\

\bottomrule
\end{tabular}
\end{center}
\end{table}

\subsubsection{More Interfering Speakers}
To determine whether the injected linguistic knowledge can mitigate the extraction challenges caused by noise, we evaluate our approach on mixture speech utterances containing one target speaker and two interfering speakers.

 As reported in Table \ref{3spk}, both  USEV and AV-Mamba with all three guidance strategies show modest performance compared to performance on two-speaker mixtures, primarily due to the zero-shot setting. Nevertheless, slight improvements were observed across all three strategies in terms of SI-SDR-i. For USEV, the false extraction rate decreased from 6.34\% to 5.10\%, and for AV-Mamba from 9.90\% to 6.86\%. It suggests that proposed strategies could alleviate the target confusion problem. 
 


 \subsubsection{Out-of-domain Test Set}
In this section, we conduct a cross-domain evaluation by evaluating the pretrained models on the LRS3 test set. Unlike VoxCeleb2~\cite{Afouras2018LRS3TEDAL}, which contains more background noise and music, the LRS3 dataset is sourced from TED videos and has a cleaner acoustic environment. The results, summarized in Table \ref{LRS3}, show that all models perform better on LRS3 in terms of signal similarity, semantic similarity, and speech intelligibility, likely due to the reduced noise interference. For USEV, the output guidance approach delivers the most significant improvement, while for AV-Mamba, the intermediate guidance method stands out. Specifically, AV-Mamba achieves an SI-SDR of 14.507 dB and a SpeechBERTScore of 0.921, demonstrating strong performance.

\begin{table*}
\caption{Offline inference results with clean visual cues on the core test set are presented. USEV serves as the baseline, and three different LLMs are evaluated: RoBERTa-base, Qwen3-0.6B, and Qwen3-4B. The evaluation includes both non-autoregressive (NAR) and autoregressive (AR) LLMs, as well as AR LLMs of varying sizes. For each model, results are reported using three distinct guidance strategies. All models are pretrained on the core training set with clean visual cues.}
\label{diff_llm}
\centering
\setlength{\tabcolsep}{6pt} 
\renewcommand{\arraystretch}{1.15} 
\begin{tabular}{
    l 
    l 
    l 
    S[table-format=2.3] 
    S[table-format=2.3] 
    S[table-format=2.3] 
    S[table-format=1.3] 
    S[table-format=1.3] 
    S[table-format=1.3] 
  }
\toprule
\multirow{2}{*}{\textbf{AV-TSE}} & \multirow{2}{*}{\textbf{LLM}} & \multirow{2}{*}{\textbf{Guidance}}
& \multicolumn{3}{c}{\textbf{Distortion (dB)}} & \multicolumn{3}{c}{\textbf{Perceptual/Intelligibility/Semantic}} \\
\cmidrule(lr){4-6} \cmidrule(lr){7-9}
& & & \textbf{SI-SDR} & \textbf{SI-SDR-i} & \textbf{SDR-i} & \textbf{PESQ} & \textbf{STOI} & \textbf{SpeechBERTScore} \\
\toprule
\multirow{9}{*}{USEV}
  & \cellcolor{lightgray}- & \cellcolor{lightgray}Baseline & \cellcolor{lightgray}10.749& \cellcolor{lightgray}10.797& \cellcolor{lightgray}11.298&\cellcolor{lightgray}2.845&\cellcolor{lightgray}0.861&\cellcolor{lightgray}0.806\\
\cline{2-9}
&\multirow{3}{*}{Roberta-base (NAR)}& Input & \textbf{11.478}& \textbf{11.525}  & \textbf{11.832}& 2.083& \textbf{0.869}&\textbf{0.813}  \\
 && Intermediate & 11.297  & 11.344& 11.652& 2.847&0.867&0.810  \\
 &&  Output &11.416 & 11.464& 11.777& \textbf{2.907}&0.868&0.812\\
 
 \cline{2-9}
&\multirow{3}{*}{Qwen3-0.6B (AR)}& Input & 11.006&11.053& 11.364& 2.814& 0.861&0.803  \\
 && Intermediate & \textbf{11.324}  &\textbf{11.371}&\textbf{11.691}&\textbf{2.855}&\textbf{0.867}&\textbf{0.810}  \\
 &&  Output & 11.283 & 11.331& 11.642& 2.850& 0.867&  0.809\\
 \cline{2-9}
&\multirow{3}{*}{Qwen3-4B (AR)}& Input & \textbf{11.840}&\textbf{11.887}& \textbf{12.196}& 2.793&\textbf{0.874}  &\textbf{0.818} \\
 && Intermediate & 11.312   &11.359 & 11.666&2.851 &0.867&0.809\\
 &&  Output & 11.536 & 11.583&11.888&\textbf{2.886}& 0.871&0.813 \\
\toprule
 
\end{tabular}
\end{table*}

\subsection{How LLM type and model size matter?}

To investigate whether different LLMs contribute guidance at varying levels, we evaluated the three proposed strategies using Roberta-base (NAR), Qwen3-0.6B (AR), and Qwen3-4B(AR) on USEV  for comparison. \footnote{Note that due to computing resource constraints, we only use USEV as the backbone for verification experiments in the following sections.}

As shown in Table \ref{diff_llm}, both AR and NAR LLMs with any type of guidance can enhance performance. Among these, Qwen3-4B achieves the most significant improvement, with an SI-SDR increase of 11.840 and a 1.091 dB gain over the baseline.

For NAR LLM, Roberta-base, we observe that intermediate guidance brings modest improvements compared to the other two guidance strategies, which might be because Roberta is pretrained with the NAR strategy. We utilized the causal architecture of Roberta with a decoder, which is not aligned with the pre-training stage.

For AR LLM Qwen3-0.6B and Qwen3-4B, similar performances were observed on intermediate guidance. For input and output guidance, Qwen3-4B is significantly better than Qwen3-0.6B. 
We hypothesize that the embeddings from the final layers of Qwen3-4B encapsulate more comprehensive utterance-level semantic information, making them more appropriate for input and output guidance strategies. The intermediate features from both Qwen3-4B and Qwen3-0.6B are likely to contain comparable syntactic-level linguistic knowledge and may result in similar performance. However, they could slightly outperform NAR LLM, RoBERTa-base, when utilizing intermediate guidance. We also observed that Qwen3-0.6B with input guidance brings modest improvements to Roberta-base and Qwen3-4B, perhaps due to the limited size of LLM.

Overall, we observe that  LLM types indeed have an impact on final performance on the same AV-TSE backbone, and a larger size LLM pre-trained with more data might provide more useful linguistic guidance for AV-TSE models.

\begin{table}
\caption{Offline inference results on core test set. All models are pre-trained on the core training set. The visual cue conditions include both impaired and clean scenarios. In each condition, both training and testing sets remain consistent. We report the SI-SDR results, as well as the SI-SDR relative improvements compared to the baseline results, denoted as `RI'.}
\label{visual_compensate}
\centering
\setlength{\tabcolsep}{6pt} 
\renewcommand{\arraystretch}{1.15} 
\begin{tabular}{
    p{0.04\textwidth} 
    p{0.05\textwidth} 
    p{0.07\textwidth} 
    p{0.07\textwidth} 
    S[table-format=2.3] 
    S[table-format=2.3] 
  }
  \toprule
  \multirow{2}{*}{\textbf{AV-TSE}} & \multirow{2}{*}{\textbf{Cue}} & \multirow{2}{*}{\textbf{LLM}} & \multirow{2}{*}{\textbf{Guidance}}
  & \multicolumn{2}{c}{\textbf{Distortion (dB)}} \\
  \cmidrule(lr){5-6}
  & & & & \textbf{SI-SDR} & \textbf{RI} \\
  \midrule
\multirow{12}{*}{USEV}&\multirow{7}{*}{Impaired}& \cellcolor{lightgray}Baseline & \cellcolor{lightgray}- &\cellcolor{lightgray}9.433& \cellcolor{lightgray}0\%\\
\cline{3-6}
 & &\multirow{3}{*}{Roberta-base} & Input & \textbf{10.683}&\textbf{13.25\%}\\
 & & & Intermediate &9.909 &5.05\% \\
 & & & Output &10.261 &8.78\% \\
 \cline{3-6}
 & & \multirow{3}{*}{Qwen3-0.6B} & Input & 9.591 &1.67\% \\
 & & & Intermediate &9.883 & 4.77\% \\
 & & & Output & 9.809 & 3.99\%\\
\cline{2-6}
 &\multirow{7}{*}{Clean} &  \cellcolor{lightgray}Baseline &  \cellcolor{lightgray}-&\cellcolor{lightgray}10.749 & \cellcolor{lightgray}0\% \\
 \cline{3-6}
 & &\multirow{3}{*}{Roberta-base} & Input & \textbf{11.478}&\textbf{6.78\%}\\
 & & & Intermediate & 11.297 &5.10\% \\
 & & & Output & 11.416& 6.21\% \\
 \cline{3-6}
 & & \multirow{3}{*}{Qwen3-0.6B} & Input & 11.053 & 2.83\%\\
 & & & Intermediate & 11.324
 &5.35\%  \\
 & & & Output & 11.283&4.97\% \\
\toprule
\end{tabular}
\end{table}

\subsection{Do LLM guidance compensate when the primary visual cue is impaired?}
In this setting, we explore whether incorporated linguistic knowledge could compensate more when the primary visual cue is impaired. To verify this, we select USEV as the AV-TSE backbone and pre-train USEV with either visual clean or visual impaired conditions. For LLM, considering computation efficiency, we select one NAR LLM, Roberta-base, and one AR LLM, Qwen3-0.6B, for exploration.

As shown in Table \ref{visual_compensate}, when using RoBERTa-base, the relative SI-SDR improvements are more pronounced under impaired visual cue conditions compared to the clean scenario. In particular, with the input guidance strategy, the improvement reaches 13.25\% under impairment, compared to 6.78\% in the clean case. This suggests that when the primary visual cue is compromised, the deep linguistic knowledge from the LLM can serve as compensation cues.  

In contrast, using Qwen3-0.6B achieves relative modest improvements on the impaired visual cue. Besides, we also note that for either visual cue is clean or impaired, using Roberta-base is more effective than using Qwen3-0.6B. To explain this, we hypothesize that Qwen3-0.6B may possess strong generative capabilities as an AR LLM. However, its generalization ability appears to be more modest compared to RoBERTa-base, particularly when it is used solely as a frozen linguistic knowledge encoder.

This approach is highly valuable in real-world applications, as scenarios with visual impairments are often more complex. However, collecting such data can be challenging. With the LLM guidance strategy, no additional data is needed—only leveraging linguistic knowledge to compensate for visual cue impairments.

\vspace{-5pt}
\begin{figure}[htbp]
 \centering
\includegraphics[scale=0.58]{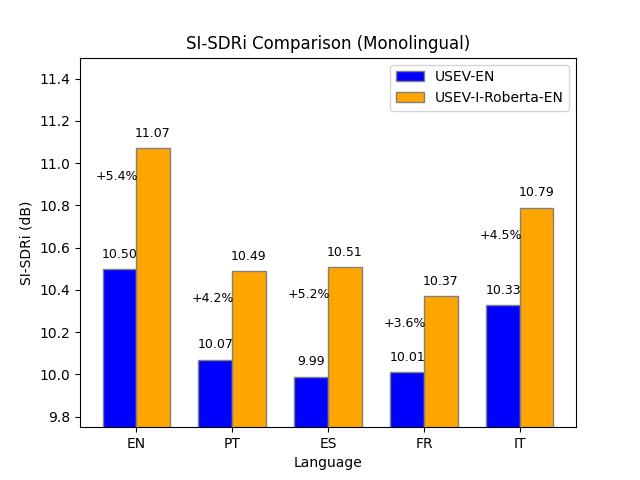}
\vspace{-5pt}
 \caption{\centering SI-SDR-i results on five monolingual test sets, USEV-EN and USEV-I-Roberta-EN denote baseline and baseline with input guidance strategy using Roberta-base. Both USEV-EN and USEV-I-Roberta-EN are trained on the English monolingual training set with clean visual cues.}
 \label{three_fusion}
\end{figure}

\subsection{
Are LLM guidance strategies language-dependent for AV-TSE, and how do they perform specifically in cross-lingual settings?}
In this section, we seek to evaluate whether the proposed LLM guidance strategies are language-dependent for AV-TSE models. Specifically, we aim to explore: (1) whether linguistic knowledge obtained from an LLM in one language can be effectively transferred to unseen languages in AV-TSE models;
(2) Given that linguistic features such as syntax differ across languages and can facilitate extraction from a multilingual mixture, we evaluate whether the advantages of linguistic knowledge are less prominent in a monolingual mixture compared to a multilingual mixture.
To explore these two questions, we select USEV as the baseline and USEV with input guidance from Roberta-base for comparison. Both models are pre-trained on the English monolingual training set with clean visual cues and tested on five monolingual test sets.

As shown in Fig.\ref{three_fusion}, we observe that (1) even unexposed to PT, ES, FR, IT, and only exposed to the English monolingual training set, USEV-EN still works on the  PT, ES, FR, IT monolingual mixtures. With LLM input guidance, USEV-I-Roberta-EN achieves consistent improvements on all five monolingual test sets. Among them, the most significant improvements were achieved on EN, with 5.4\% relative improvements. This is reasonable because the model is trained exclusively on English data; even with the highest baseline performance, the relative improvements remain the greatest, further supporting the effectiveness of the proposed LLM guidance strategy. For unseen languages, ES achieves second-best, with 5.2\% relative improvements, perhaps because the baseline results are lowest. These results suggest that linguistic
knowledge obtained from an LLM in one language can be
effectively transferred to unseen languages. It is practical for  AV-TSE on extremely low-resource languages, where collecting a corresponding multimodal dataset is challenging. (2) Compared to the results on the core test set, both USEV and USEV-I-Roberta-EN achieve modest performance on monolingual test sets. Though core training and test sets contain both multi-lingual and mono-lingual mixtures, the extraction is still relatively easier. On the core test set, in terms of SI-SDR-i, USEV achieves 10.749, and USEV-I-Roberta achieves 11.525, with 7.16\% relative improvements, higher than 5.4\% improvements achieved on EN monolingual. It suggests that the gain of linguistic knowledge on extraction performance is reduced, which aligns with our hypothesis.

\begin{table}
\centering
\label{attention}
\caption{SI-SDR comparison results of using Cross-Attention and Gated Attention. `USEV-I-Ro' denotes USEV with input guidance from Roberta.}
\setlength{\tabcolsep}{6pt} 
\renewcommand{\arraystretch}{1.15} %
\begin{tabular}{lll}
\toprule
\textbf{Model} & \textbf{Type} & \textbf{SI-SDR} \\
\toprule
\multirow{2}{*}{USEV-I-Ro} &+ Gated-attention & 11.478 \\
 &+ Cross-attention & 11.101 \\
\toprule
\end{tabular}
\end{table}

\begin{figure*}[htbp]
 \centering
\includegraphics[scale=0.35]{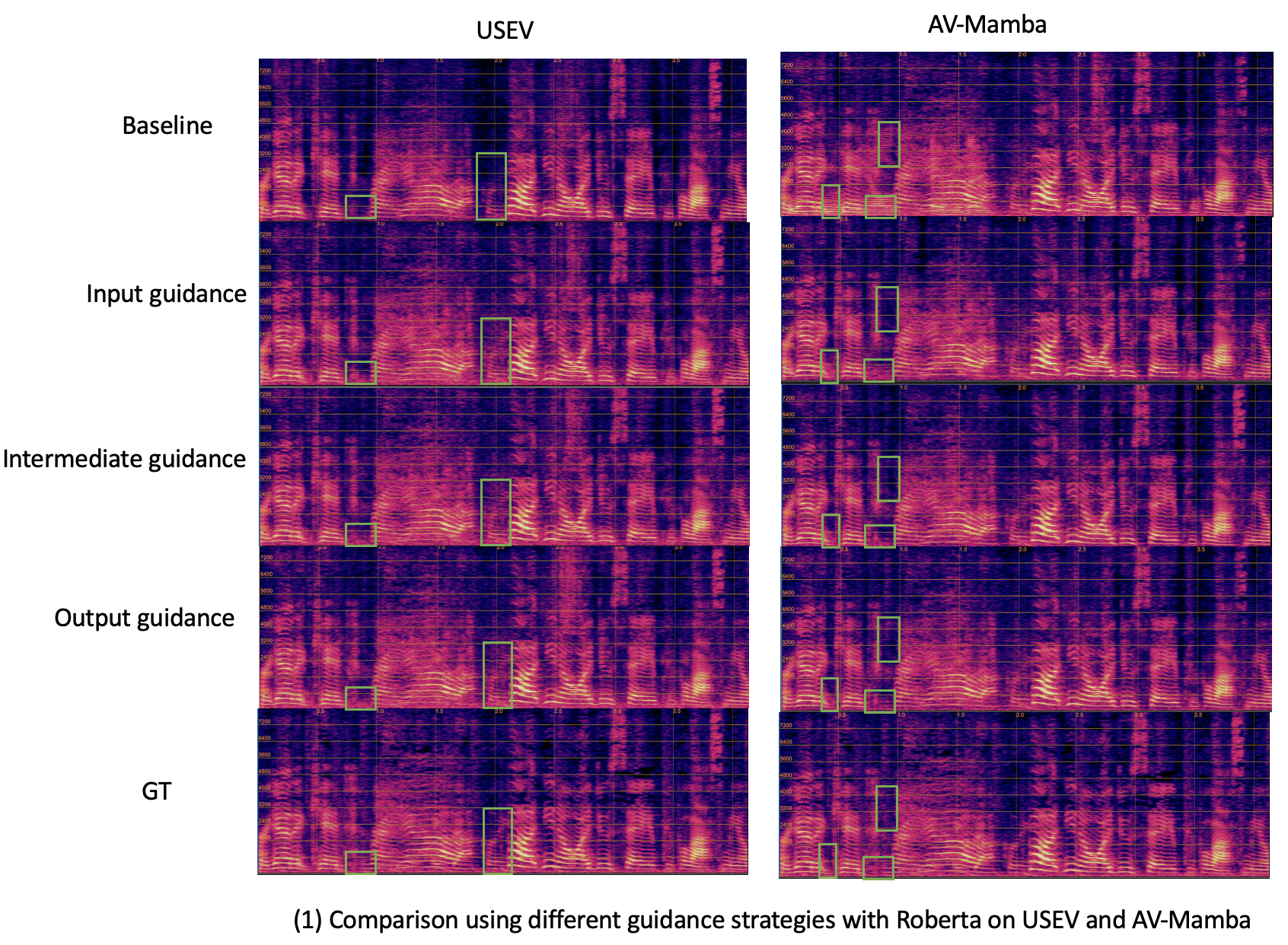}
  \vspace{-10pt}
 \caption{\centering Mel-spectrograms of USEV and AV-Mamba extraction results using three proposed strategies with Roberta.}
  \vspace{-10pt}
 \label{case1}
\end{figure*}

\begin{figure*}[htbp]
 \centering
\includegraphics[scale=0.4]{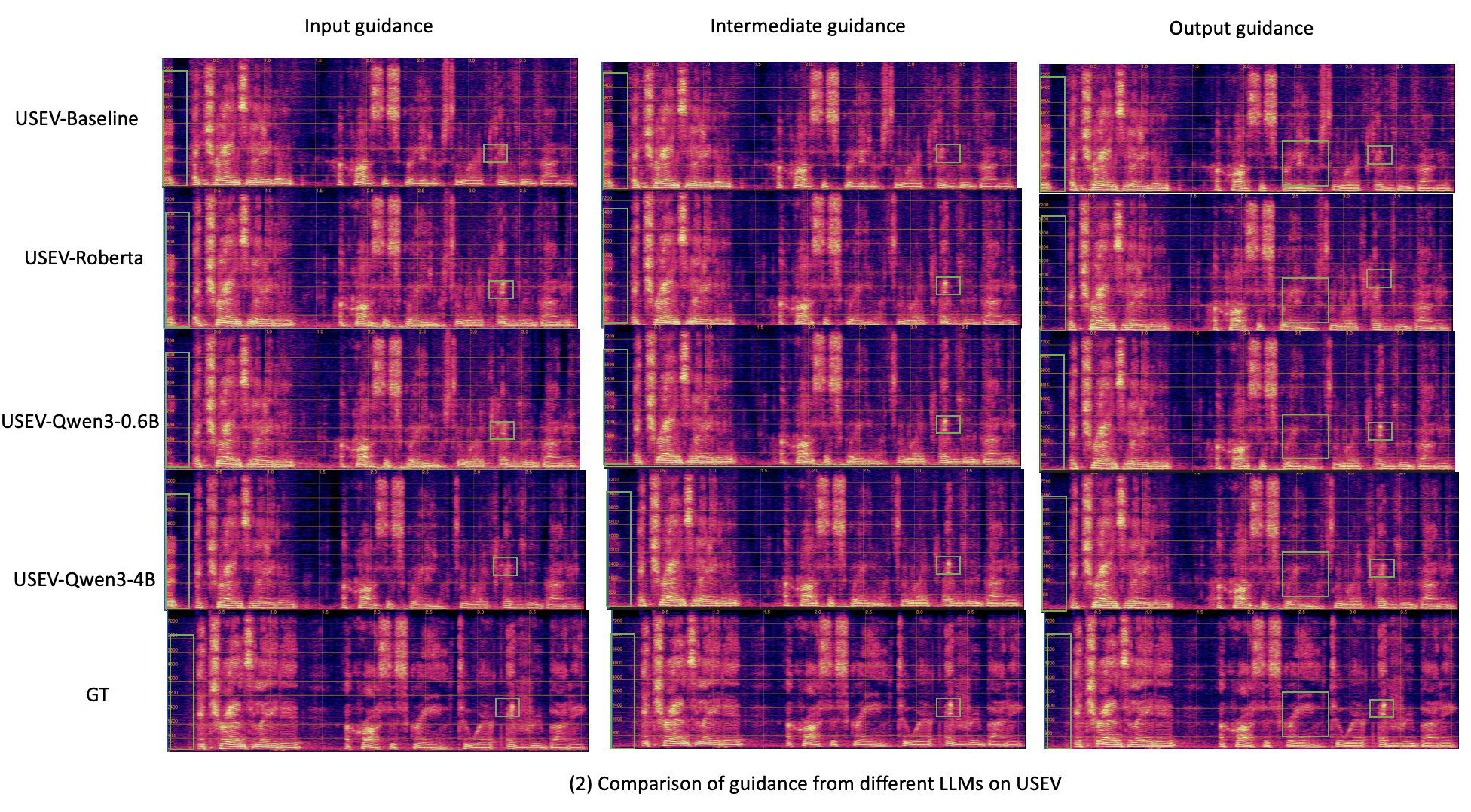}
  \vspace{-10pt}
 \caption{\centering Mel-spectrograms of USEV extraction results with the combination of three proposed strategies with Roberta, Qwen3-0.6B, and Qwen3-4B.}
  \vspace{-10pt}
 \label{case2}
\end{figure*}

\begin{figure*}[htbp]
 \centering
\includegraphics[scale=0.45]{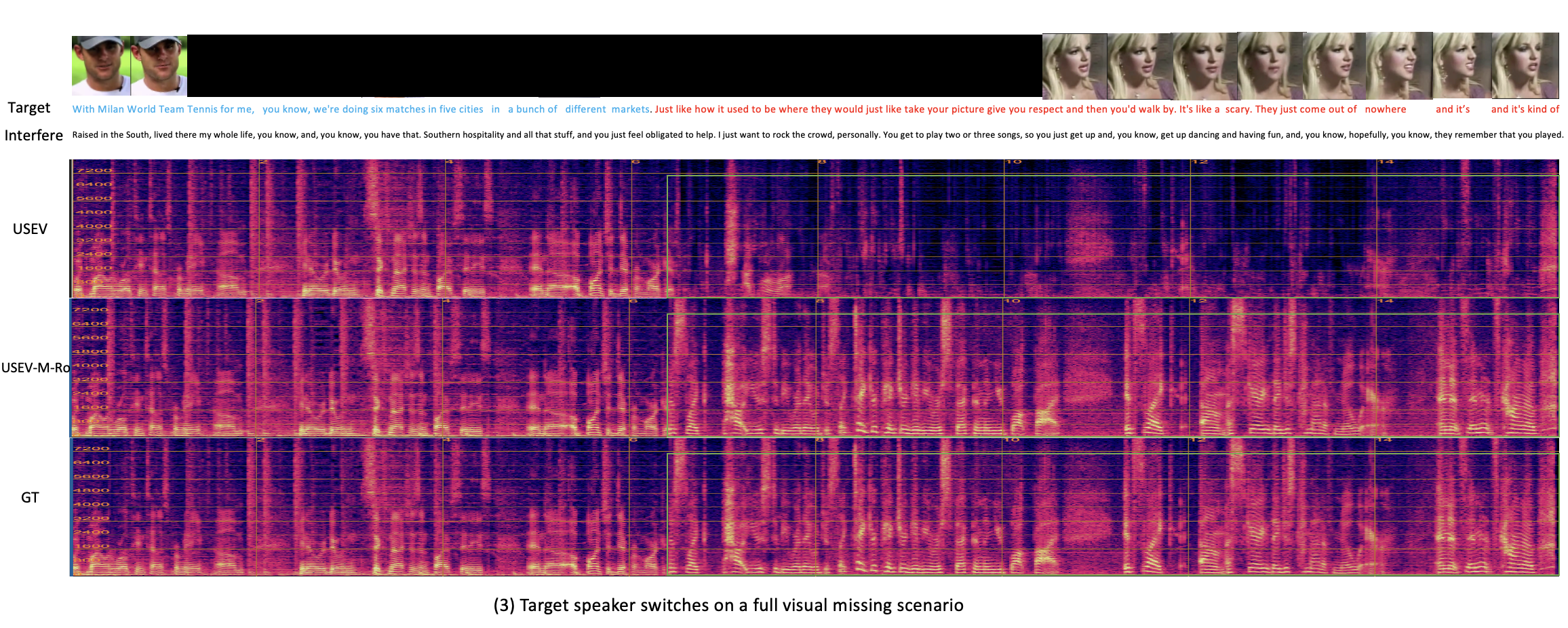}
  \vspace{-15pt}
 \caption{\centering  Impaired target visual cues, GT transcriptions of two target speakers (blue and red text) and one interfering speaker (black text), and mel-spectrograms on the target speaker switches scenario. USEV with intermediate guidance from Roberta is denoted as `USEV-M-Ro'.}
   \vspace{-10pt}
 \label{case3}
\end{figure*}

\subsection{Ablation Study}
For input guidance, we explored using cross-attention or gated attention between the LLM embedding and the mixture speech embedding. As shown in Table \ref{attention}, we found that using gated attention achieves better results in terms of SI-SDR, so we keep this experimental setting. We assume gated attention retrieves the most vital target speech content from the mixture signal, which is more useful for extraction with linguistic prior, as we introduced in the method section.

For intermediate guidance, we conducted preliminary experiments and found that increasing the guidance factor $\alpha$ from 0.1 to larger values may lead to an early stop and will result in negative performance gains. We assume that too much acoustic information might lead to a negative effect on LLM's NTP ability, so we keep $\alpha$ to 0.1 in all experimental settings.


\subsection{Case Study on Improvements on Core Test Set}
This section provides mel-spectrogram visualizations of extracted speech utterances to illustrate the performance improvements from our proposed strategies.
\subsubsection{Input vs Middle vs Output}
As shown in Although the AV-Mamba baseline exhibits superior overall performance on the core test set, we highlight an example where the USEV baseline achieves marginally better results, shown in Fig.\ref{case1}. While both baselines produce relatively accurate outputs, all three proposed strategies demonstrate improvements, with enhancements in low-frequency regions specifically indicated by green rectangles.

\subsubsection{Roberta vs Qwen3-0.6B vs Qwen3-4B}
Fig.\ref{case2} compares the performance of all combinations of three LLM types and three guidance strategies, using a USEV baseline. We observe two primary improvements within the same mixture speech scenario: (1) At the utterance's beginning, where silence is expected, the baseline, as well as input and intermediate guidance using any LLM, falsely extract audio. Only output guidance consistently avoids this artifact across all three LLMs. (2) A significant low-frequency component is present in the second half of the extracted utterance, which the baseline fails to capture. Even when using the same LLM (e.g., Qwen3-4B), input guidance misses this component, while intermediate and output guidance successfully extract it. These results suggest that neither LLM size nor guidance strategy alone guarantees success; rather, the effectiveness depends on their combination.
\subsubsection{Target Speaker Switching}
Fig.\ref{case3} illustrates a particularly challenging target speaker switching scenario where visual cues are severely compromised across most frames, including the critical switching point. The blue text indicates the first target speaker's utterance content, and the red text indicates the second. With the visual cue entirely absent from 1 to 10 seconds, completely encompassing the switching point, the baseline failed to extract the second target speaker's speech, resulting in near silence evident in the spectrogram. It is understandable, given its reliance on visual cues. In contrast, USEV with RoBERTa-based input guidance, pre-trained on a visual cue impaired dataset, effectively extracts both target speakers' speech, achieving an SI-SDR exceeding 19dB. This demonstrates the effectiveness of linguistic cues in compensating for severe visual impairment, especially in extremely challenging scenarios.

\section{Conclusion}
This study presents three novel LLM guidance strategies for enhancing AV-TSE models by incorporating linguistic knowledge at the training phase. These strategies were rigorously validated using two baselines, including USEV and AV-Mamba (a pioneering application of Mamba in AV-TSE). Extensive experiments on different LLMs (RoBERTa-base, Qwen3-0.6B, Qwen3-4B), and challenging scenarios, including visual cue impaired, different monolingual mixtures, increased interfering speakers, and target speaker switching, demonstrated the robustness of the proposed three strategies. In the future, we will explore more efficient knowledge injection methods to further improve AV-TSE models.

\bibliographystyle{IEEEtran}
\bibliography{refs} 

@article{Hong2023Watch,
  title={Watch or Listen: Robust Audio-Visual Speech Recognition with Visual Corruption Modeling and Reliability Scoring},
  author={Joanna Hong and Minsu Kim and Jeong Yun Choi and Yong Man Ro},
  journal={2023 CVPR}
}

@article{gpt4o,
  title={Gpt-4o system card},
  author={Hurst, Aaron and Lerer, Adam and Goucher, Adam P and Perelman, Adam and Ramesh, Aditya and Clark, Aidan and Ostrow, AJ and Welihinda, Akila and Hayes, Alan and Radford, Alec and others},
  journal={arXiv preprint arXiv:2410.21276},
  year={2024}
}

@article{thimonier2025emosllm,
  title={EmoSLLM: Parameter-Efficient Adaptation of LLMs for Speech Emotion Recognition},
  author={Thimonier, Hugo and Perzo, Antony and Seguier, Renaud},
  journal={arXiv preprint arXiv:2508.14130},
  year={2025}
}

@inproceedings{yu2024repa,
    title={Representation Alignment for Generation: Training Diffusion Transformers Is Easier Than You Think},
    author={Sihyun Yu and Sangkyung Kwak and Huiwon Jang and Jongheon Jeong and Jonathan Huang and Jinwoo Shin and Saining Xie},
    year={2025},
    booktitle={International Conference on Learning Representations (ICLR)},
}

@article{MultiModalCorrelation,
author = {Xiong, Junwen and Zhou, Yu and Zhang, Peng and Xie, Lei and Huang, Wei and Zha, Yufei},
year = {2022},
month = {01},
pages = {1-14},
title = {Look\&listen: Multi-Modal Correlation Learning for Active Speaker Detection and Speech Enhancement},
volume = {PP},
journal = {IEEE Transactions on Multimedia (TMM)},
doi = {10.1109/TMM.2022.3199109}
}

@article{yeo2024akvsr,
  title={Akvsr: Audio knowledge empowered visual speech recognition by compressing audio knowledge of a pretrained model},
  author={Yeo, Jeong Hun and Kim, Minsu and Choi, Jeongsoo and Kim, Dae Hoe and Ro, Yong Man},
  journal={IEEE Transactions on Multimedia (TMM)},
  volume={26},
  pages={6462--6474},
  year={2024},
  publisher={IEEE}
}

@INPROCEEDINGS{misp,
  author={Chen, Hang and Wu, Shilong and Wang, Chenxi and Du, Jun and Lee, Chin-Hui and Siniscalchi, Sabato Marco and Watanabe, Shinji and Chen, Jingdong and Scharenborg, Odette and Wang, Zhong-Qiu and Yin, Bao-Cai and Pan, Jia},
  booktitle={2024 IEEE ICASSPW}, 
  title={Summary on the Multimodal Information-Based Speech Processing (MISP) 2023 Challenge}, 
  year={2024},
}

@INPROCEEDINGS{HRI,
  author={Goetzee, Sander and Mihhailov, Konstantin and Van De Laar, Roel and Baraka, Kim and Hindriks, Koen V.},
  booktitle={2024 33rd IEEE International Conference on Robot and Human Interactive Communication (ROMAN)}, 
  title={Audio-Visual Speech Recognition for Human-Robot Interaction: a Feasibility Study}, 
  year={2024},
  volume={},
  number={},
  pages={930-935},
  doi={10.1109/RO-MAN60168.2024.10731139}}

@misc{qwen3technicalreport,
      title={Qwen3 Technical Report}, 
      author={Qwen Team},
      year={2025},
      eprint={2505.09388},
      archivePrefix={arXiv},
      primaryClass={cs.CL},
      url={https://arxiv.org/abs/2505.09388}, 
}

@article{Liu2019RoBERTaAR,
  title={RoBERTa: A Robustly Optimized BERT Pretraining Approach},
  author={Yinhan Liu and Myle Ott and Naman Goyal and Jingfei Du and Mandar Joshi and Danqi Chen and Omer Levy and Mike Lewis and Luke Zettlemoyer and Veselin Stoyanov},
  journal={ArXiv},
  year={2019},
}

@INPROCEEDINGS{ijcnn,
  author={\vspace{0mm}Wu, Wenxuan and Chen, Xueyuan and Wu, Xixin and Li, Haizhou and Meng, Helen},
  booktitle={IJCNN 2024}, 
  title={Target Speech Extraction with Pre-trained AV-HuBERT and Mask-And-Recover Strategy}}

@INPROCEEDINGS{mengicassp2025,
  author={Meng, Lingwei and Hu, Shujie and Kang, Jiawen and Li, Zhaoqing and Wang, Yuejiao and Wu, Wenxuan and Wu, Xixin and Liu, Xunying and Meng, Helen},
  booktitle={ICASSP 2025}, 
  title={Large Language Model Can Transcribe Speech in Multi-Talker Scenarios with Versatile Instructions}, 
}

@inproceedings{Cappellazzo2024LargeLM,
  title={Large Language Models Are Strong Audio-Visual Speech Recognition Learners},
  author={Umberto Cappellazzo and Minsu Kim and Honglie Chen and Pingchuan Ma and Stavros Petridis and Daniele Falavigna and Alessio Brutti and Maja Pantic},
  booktitle={IEEE International Conference on Acoustics, Speech, and Signal Processing},
  year={2024},
 
}

@inproceedings{whisper-llama,
    title = "Whispering {LL}a{MA}: A Cross-Modal Generative Error Correction Framework for Speech Recognition",
    author = "Radhakrishnan, Srijith  and
      Yang, Chao-Han Huck  and
      Khan, Sumeer Ahmad  and
      Kumar, Rohit  and
      Kiani, Narsis A.  and
      Gomez-Cabrero, David  and
      Tegner, Jesper N.",
    editor = "Bouamor, Houda  and
      Pino, Juan  and
      Bali, Kalika",
    booktitle = "Proceedings of the 2023 Conference on Empirical Methods in Natural Language Processing (EMNLP)",
    month = dec,
    year = "2023",
}

@article{shi2024lmfusion,
  title={LMFusion: Adapting Pretrained Language Models for Multimodal Generation},
  author={Shi, Weijia and Han, Xiaochuang and Zhou, Chunting and Liang, Weixin and Lin, Xi Victoria and Zettlemoyer, Luke and Yu, Lili},
  journal={arXiv preprint arXiv:2412.15188},
  year={2024}
}

@unknown{DiffDSR,
author = {Chen, Xueyuan and Yang, Dongchao and Wu, Wenxuan and Wu, Minglin and Xu, Jing and Wu, Xixin and Wu, Zhiyong and Meng, Helen},
year = {2025},
month = {05},
pages = {},
title = {DiffDSR: Dysarthric Speech Reconstruction Using Latent Diffusion Model},
booktitle = {Interspeech 2025}
}

@inproceedings{LLMttst5,
author = {Neekhara, Paarth and Hussain, Shehzeen and Ghosh, Subhankar and Li, Jason and Ginsburg, Boris},
year = {2024},
month = {09},
pages = {3425-3429},
title = {Improving Robustness of LLM-based Speech Synthesis by Learning Monotonic Alignment},
booktitle = {Interspeech 2024}
}

@article{Afouras2018LRS3TEDAL,
  title={LRS3-TED: a large-scale dataset for visual speech recognition},
  author={Triantafyllos Afouras and Joon Son Chung and Andrew Zisserman},
  journal={ArXiv},
  year={2018},
}

@inproceedings{avsepchain,
author = {Mu, Zhaoxi and Yang, Xinyu},
title = {Separate in the speech chain: cross-modal conditional audio-visual target speech extraction},
booktitle = {IJCAI 2024}}

@inproceedings{saeki2024speechbertscore,
  title={{SpeechBERTScore}: Reference-Aware Automatic Evaluation of Speech Generation Leveraging NLP Evaluation Metrics},
  author={Takaaki Saeki and Soumi Maiti and Shinnosuke Takamichi and Shinji Watanabe and Hiroshi Saruwatari},
  booktitle={Interspeech 2024},
}

@inproceedings{kim25q_interspeech,
  title     = {{Mamba-based Hybrid Model for Speech Enhancement}},
  author    = {Se-Ha Kim and Tae-Gyeong Kim and Chang-Jae Chun},
  year      = {2025},
  booktitle = {{Interspeech 2025}},
  pages     = {5163--5167},
  doi       = {10.21437/Interspeech.2025-1476},
  issn      = {2958-1796},
}

@inproceedings{jiang2025speech,
  title={Speech slytherin: Examining the performance and efficiency of mamba for speech separation, recognition, and synthesis},
  author={Jiang, Xilin and Li, Yinghao Aaron and Florea, Adrian Nicolas and Han, Cong and Mesgarani, Nima},
  booktitle={ICASSP 2025-2025 IEEE International Conference on Acoustics, Speech and Signal Processing (ICASSP)},
  pages={1--5},
  year={2025},
  organization={IEEE}
}

@InProceedings{Masuyama2024SLT,
  author    =  {Masuyama, Yoshiki and Miyazaki, Koichi and Murata, Masato},
  title     =  {Mamba-based Decoder-Only Approach with Bidirectional Speech Modeling for Speech Recognition},
  booktitle =  {Proc. IEEE Spoken Language Technology Workshop (SLT)},
  year      =  2024,
  month     =  dec
}

@inproceedings{chao2024investigation,
  title={An investigation of incorporating mamba for speech enhancement},
  author={Chao, Rong and Cheng, Wen-Huang and La Quatra, Moreno and Siniscalchi, Sabato Marco and Yang, Chao-Han Huck and Fu, Szu-Wei and Tsao, Yu},
  booktitle={2024 IEEE Spoken Language Technology Workshop (SLT)},
  pages={302--308},
  year={2024},
  organization={IEEE}
}

@inproceedings{huo2025beyond,
  title={Beyond speaker identity: Text guided target speech extraction},
  author={Huo, Mingyue and Jain, Abhinav and Huynh, Cong Phuoc and Kong, Fanjie and Wang, Pichao and Liu, Zhu and Bhat, Vimal},
  booktitle={ICASSP 2025},
  pages={1--5},
  year={2025},
  organization={IEEE}
}

@inproceedings{kim2025contextual,
  title={Contextual Speech Extraction: Leveraging Textual History as an Implicit Cue for Target Speech Extraction},
  author={Kim, Minsu and Mira, Rodrigo and Chen, Honglie and Petridis, Stavros and Pantic, Maja},
  booktitle={ICASSP 2025},
  pages={1--5},
  year={2025},
  organization={IEEE}
}

@ARTICLE{LISTEN-CHAT-REMIX,
  author={Jiang, Xilin and Han, Cong and Li, Yinghao Aaron and Mesgarani, Nima},
  journal={IEEE Journal of Selected Topics in Signal Processing}, 
  title={Listen, Chat, and Remix: Text-Guided Soundscape Remixing for Enhanced Auditory Experience}, 
  year={2025},
  volume={19},
  number={4},
  pages={635-645},
  doi={10.1109/JSTSP.2025.3570103}}

@article{li2025memo,
  title={MeMo: Attentional Momentum for Real-time Audio-visual Speaker Extraction under Impaired Visual Conditions},
  author={Li, Junjie and Wu, Wenxuan and Wang, Shuai and Pan, Zexu and Lee, Kong Aik and Meng, Helen and Li, Haizhou},
  journal={arXiv preprint arXiv:2507.15294},
  year={2025}
}

@article{wu2025c,
  title={$ C^2$ AV-TSE: Context and Confidence-aware Audio Visual Target Speaker Extraction},
  author={Wu, Wenxuan and Chen, Xueyuan and Wang, Shuai and Wang, Jiadong and Meng, Lingwei and Wu, Xixin and Meng, Helen and Li, Haizhou},
  journal={IEEE Journal of Selected Topics in Signal Processing},
  year={2025},
  publisher={IEEE}
}

@inproceedings{wu2025incorporating,
author = {Wu, Wenxuan and Wang, Shuai and Wu, Xixin and Meng, Helen and Li, Haizhou},
title = {Incorporating Linguistic Constraints from External Knowledge Source for Audio-Visual Target Speech Extraction},
booktitle = {Interspeech 2025},
}

@article{brain,
  title={Robust speech processing in human auditory cortex},
  author={Mesgarani, Nima},
  journal={The Journal of the Acoustical Society of America},
  volume={143},
  number={3},
  pages={1744--1744},
  year={2018},
  publisher={Acoustical Society of America}
}

@article{select-listen,
author = {Pan, Zexu and Tao, Ruijie and Xu, Chenglin and Li, Haizhou},
year = {2022},
month = {01},
pages = {1-1},
title = {Selective Listening by Synchronizing Speech With Lips},
volume = {30},
journal = {IEEE/ACM Transactions on Audio, Speech, and Language Processing}
}

@article{muse,
  
  author={\vspace{0mm}Z. Pan and Ruijie Tao and Chenglin Xu and Haizhou Li},
 title={Muse: Multi-Modal Target Speaker Extraction with Visual Cues},
  journal={ICASSP 2021},
}

@unknown{hsieh2024multimodal,
author = {Hsieh, Tsun-An and Choi, Heeyoul "Henry and Kim, Minje},
year = {2024},
month = {06},
title = {Multimodal Representation Loss Between Timed Text and Audio for Regularized Speech Separation},
 booktitle={Interspeech 2024},
}

@article{hung2025linguistic,
  title={Linguistic Knowledge Transfer Learning for Speech Enhancement},
  author={Hung, Kuo-Hsuan and Lu, Xugang and Fu, Szu-Wei and Tseng, Huan-Hsin and Lin, Hsin-Yi and Lin, Chii-Wann and Tsao, Yu},
  journal={arXiv preprint arXiv:2503.07078},
  year={2025}
}

@inproceedings{whisper,
author = {Radford, Alec and Kim, Jong Wook and Xu, Tao and Brockman, Greg and McLeavey, Christine and Sutskever, Ilya},
title = {Robust speech recognition via large-scale weak supervision},
year = {2023},
booktitle = {Proceedings of the 40th International Conference on Machine Learning (ICML)},
articleno = {1182},
numpages = {27},
location = {Honolulu, Hawaii, USA},

}

@inproceedings{DistillationASR,
author = {Lee, Jeehye and Seo, Hyeji},
year = {2024},
month = {09},
pages = {2890-2894},
title = {Online Knowledge Distillation of Decoder-Only Large Language Models for Efficient Speech Recognition},
 booktitle={Interspeech 2024}, 

}

@INPROCEEDINGS{ImagineNET,
  author={Pan, Zexu and Wang, Wupeng and Borsdorf, Marvin and Li, Haizhou},
  booktitle={ICASSP 2023}, 
  title={ImagineNet: Target Speaker Extraction with Intermittent Visual Cue Through Embedding Inpainting}, 
}

@InProceedings{voxceleb2,
              author       = "Chung, J.~S. and Nagrani, A. and Zisserman, A.",
              title        = "VoxCeleb2: Deep Speaker Recognition",
              booktitle    = "INTERSPEECH",
              year         = "2018",
            }

@ARTICLE{deepavsr,
  author={Afouras, Triantafyllos and Chung, Joon Son and Senior, Andrew and Vinyals, Oriol and Zisserman, Andrew},
  journal={IEEE Transactions on Pattern Analysis and Machine Intelligence}, 
  title={Deep Audio-Visual Speech Recognition}, 
  year={2022},
  volume={44},
  number={12},
  pages={8717-8727},
}

@ARTICLE{tse_summary,
       author = {{Zmolikova}, Katerina and {Delcroix}, Marc and {Ochiai}, Tsubasa and {Kinoshita}, Keisuke and {{\v{C}}ernock{\'y}}, Jan and {Yu}, Dong},
        title = "{Neural Target Speech Extraction: An overview}",
      journal = {IEEE Signal Processing Magazine},
         year = 2023,
        month = may,
       volume = {40},
       number = {3},
        pages = {8-29},
         
archivePrefix = {arXiv},
       eprint = {2301.13341},
 primaryClass = {eess.AS}
}

@ARTICLE{text,
       author = {{Ohishi}, Yasunori and {Delcroix}, Marc and {Ochiai}, Tsubasa and {Araki}, Shoko and {Takeuchi}, Daiki and {Niizumi}, Daisuke and {Kimura}, Akisato and {Harada}, Noboru and {Kashino}, Kunio},
        title = "{ConceptBeam: Concept Driven Target Speech Extraction}",
      journal = {arXiv e-prints},
         year = 2022,
        month = jul,
          eid = {arXiv:2207.11964},
        pages = {arXiv:2207.11964},
    
archivePrefix = {arXiv},
       eprint = {2207.11964},
 primaryClass = {eess.AS}
}

@ARTICLE{pesq,
       author = {{Basterrech}, Sebasti{\'a}n and {Rubino}, Gerardo and {Varela}, Mart{\'\i}n},
        title = "{Single-sided Real-time PESQ Score Estimation}",
      journal = {arXiv e-prints},
         year = 2012,
        month = dec,
          eid = {arXiv:1212.6350},
        pages = {arXiv:1212.6350},
         
archivePrefix = {arXiv},
       eprint = {1212.6350},
 primaryClass = {cs.SD}
}

@article{stoi,
  title={An Algorithm for Intelligibility Prediction of Time–Frequency Weighted Noisy Speech},
  author={Cees H. Taal and Richard Christian Hendriks and Richard Heusdens and Jesper Rindom Jensen},
  journal={IEEE TASLP 2011},
}

@article{lin2025bridging,
  title={Bridging The Multi-Modality Gaps of Audio, Visual and Linguistic for Speech Enhancement},
  author={Lin, Meng-Ping and Hou, Jen-Cheng and Chen, Chia-Wei and Chien, Shao-Yi and Chen, Jun-Cheng and Lu, Xugang and Tsao, Yu},
  journal={arXiv preprint arXiv:2501.13375},
  year={2025}
}

@article{usev,
author = {Pan, Zexu and Ge, Meng and Li, Haizhou},
title = {USEV: Universal Speaker Extraction With Visual Cue},
year = {2022},
issue_date = {2022},
publisher = {IEEE Press},
volume = {30},
issn = {2329-9290},
journal = {IEEE/ACM Trans. Audio, Speech and Lang. Proc.},
month = {sep},
pages = {3032–3045},
numpages = {14}
}

@article{SDR,
  title={SDR – Half-baked or Well Done?},
  author={Jonathan Le Roux and Scott Wisdom and Hakan Erdogan and John R. Hershey},
  journal={ICASSP 2019},
}

@INPROCEEDINGS{av-sepformer,
  author={Lin, Jiuxin and Cai, Xinyu and Dinkel, Heinrich and Chen, Jun and Yan, Zhiyong and Wang, Yongqing and Zhang, Junbo and Wu, Zhiyong and Wang, Yujun and Meng, Helen},
  booktitle={ICASSP 2023}, 
  title={Av-Sepformer: Cross-Attention Sepformer for Audio-Visual Target Speaker Extraction}}

@inproceedings{luo2020dual,
  title={Dual-path rnn: efficient long sequence modeling for time-domain single-channel speech separation},
  author={Luo, Yi and Chen, Zhuo and Yoshioka, Takuya},
  booktitle={ICASSP 2020-2020 IEEE International Conference on Acoustics, Speech and Signal Processing (ICASSP)},
  pages={46--50},
  year={2020},
  organization={IEEE}
}

@article{wu2024target_cvpr,
  title={AVHuMAR: Audio-Visual Target Speech Extraction with Pre-trained AV-HuBERT and Mask-And-Recover Strategy},
  author={Wu, Wenxuan and Chen, Xueyuan and Wu, Xixin and Li, Haizhou and Meng, Helen},
  journal={CVPR 2024 Sight and Sound Workshop},
}

\end{document}